\documentclass[a4paper,conference]{IEEEtran}
\usepackage{graphicx}
\usepackage{color}
\usepackage[colorlinks=true]{hyperref}

\begin{document}

\title{In Search of $H$-theorem for \\
Ulam's Redistribution Problem}

\author{
\authorblockN{Sergey Apenko}
\authorblockA{
I. E. Tamm Theory Dept.\\
P. N. Lebedev Physical Institute\\
Email: apenko@lpi.ru}
}

\maketitle

\begin{abstract}
We discuss the possibility of deriving an $H$-theorem for the nonlinear discrete time evolution known as Ulam's redistribution of energy problem. In this model particles are paired at random and then their total energy is redistributed between them according to some probability law. It appears possible to obtain the proper $H$-function which always increases during the relaxation only for a special set of redistribution laws, given by  symmetric beta distributions. This $H$-function differs from the usual entropy by an additional term that vanishes only for the uniform redistribution law. But for arbitrary redistribution the evolution has some features of relaxation to a non-equilibrium steady state and the $H$-function is still unknown.
\end{abstract}

\IEEEpeerreviewmaketitle

\section{Introduction}
Ulam's redistribution of energy problem was  introduced in a paper by  Blackwell and Mauldin \cite{U} shortly after Stanislaw Ulam had passed away and was formulated as follows: ``Consider a vast number of particles and let us
redistribute the energy of these particles... First, pair the particles at random.
Second, for each pair, redistribute the total energy of the pair between these
particles according to some given fixed probability law of redistribution...'' Ulam had suggested that after many iterations of this process the energy distribution should finally converge to some fixed `equilibrium' distribution and this conjecture was proved in Ref. \cite{U}. 

Very similar models were introduced recently in an economic context as random market models, which assume that economic transaction occur by binary `collisions' between agents who exchange money in the
same way as particles in a gas exchange their energy. In a series of papers L\'{o}pez-Ruiz and his colleagues  \cite{gl} have proposed a discrete time nonlinear evolution equation for such a process in terms of a distribution function $p(x)$  ($x>0$)  when the redistribution law was uniform. An obvious generalization of their equation  is
\begin{equation}\label{gl}
  p'(x)=\int_0^{\infty}\int_0^{\infty}dudv \,W(x; u+v)p(u)p(v).
\end{equation}
This equation shows how the distribution function transforms on each step of iterations when  $p(x)\rightarrow p'(x)$ and $W(x; u+v)$ is the probability density of transitions $u,v\rightarrow x,u+v-x$. We also assume that  $\int xW(x;u+v)dx=(u+v)/2$ so that that the mean energy $\langle x\rangle$ is conserved. For uniform redistribution we have $W(x,u+v)=1/(u+v)$ for $u+v>x$ and zero otherwise. In this case Eq. (\ref{gl}) coincides with the one used in \cite{gl} while for nonuniform redistribution laws this evolution is close to `directed' random market models \cite{dir}.

Since physically this evolution based on binary collisions is similar in spirit to the one described by the Boltzmann equation one might expect some kind of $H$-theorem to be valid here. And indeed, for the uniform redistribution law the Boltzmann entropy
$$
 S(p)=-\int dxp(x)\ln p(x)
$$
always increases \cite{gl,A} while $p(x)$ tends to the equilibrium distribution  $p_0=\lambda\exp(-\lambda x)$, where $1/\lambda=\langle x\rangle$. 

However for an arbitrary $W(x; u+v)$ one cannot expect that the entropy $S(p)$ always grows. Thus a question arises what is the proper $H$-function (if it exists)   that is monotone during the relaxation for a general redistribution law.

\section{Two particle evolution}
Our search for the $H$-theorem is based on the approach of \cite{A} which transforms initial nonlinear problem into a linear one (supplemented by some projection operation) by introducing a two particle distribution function $f(x,y)$. After a collision and redistribution of energy $f(x,y)\rightarrow f'(x,y)$ and it is possible to write down  a simple equation describing this evolution. If the redistribution law depends only on the fraction of the total energy each particle acquires, then
$$
 W(x; u+v)=\frac{1}{u+v}D\left(\frac{x}{u+v}\right), \, u+v>x
$$
and zero for $u+v<x$, where $D(t)>0$ is a normalized distribution on $t\in[0,1]$ symmetric under $t\rightarrow 1-t$, and
\begin{equation}\label{lin}
  f'(x,y)=D\left(\frac{x}{x+y}\right)\int_0^1 d\xi\,f(\xi(x+y),(1-\xi)(x+y)),
\end{equation}
This is a linear transformation and it conserves
positivity of $f(x,y)$, its norm  and the mean `energy' $\langle x+y\rangle$.

The advantage of Eq. (\ref{lin}) is that this is a linear evolution for which the monotone function can be easily constructed. Normally it is the relative entropy with respect to a stationary state that monotonically decreases.

It should be noted here that Eq. (\ref{lin}) alone does not describe correctly the
evolution of the two-particle probability distribution in Ulam's problem. It takes
into account only collisions within fixed pairs of particles while the true evolution
includes also new random pairings of particles at each step, not accounted for in (\ref{lin}). 

The new pairing may be described as a kind of `reduction' of $f'(x,y)$ back to a factorized form
\begin{equation}\label{prime}
 f'(x,y)\rightarrow p'(x)p'(y), \quad  p'(x)\equiv\int_0^{\infty} dy f'(x,y).
\end{equation}
If we define $p'(x)$ according to Eq. (\ref{prime}) and combine it with Eq. (\ref{lin}) where  $f(x,y)=p(x)p(y)$ this will give us our original nonlinear equation (\ref{gl}). Thus each step of the nonlinear evolution may be decomposed in two, in terms of $f(x,y)$. The first one is a linear evolution (\ref{lin}) with factorized  initial condition and the second one is the reduction (\ref{prime}).

Then using some general information theory inequalities and following the same line of reasoning as in Ref. \cite{A} we finally obtain our main inequality
\begin{eqnarray}\label{main}
 S(p')+ \frac{1}{2}\int_0^{\infty} dxdy\, f'(x,y)\ln\left[D\left(\frac{x}{x+y}\right)\right]\geq \nonumber\\
  \geq S(p)+\frac{1}{2}\int_0^{\infty} dxdy\, p(x)p(y)\ln\left[D\left(\frac{x}{x+y}\right)\right]
\end{eqnarray}
Unfortunately $f'(x,y)$ cannot be expressed in terms of $p'$, therefore for arbitrary $D$  we cannot derive any $H$-theorem from Eq. (\ref{main}). There is one important case, however, when this is still possible.

\section{$H$-theorem for beta redistribution law}

Let us now take the redistribution law in the form of symmetric beta distribution
$$
D(t)=Ct^{a-1}(1-t)^{a-1}, \quad t\in [0,1]
$$
where $C$ is a normalization constant and $a>0$ is a parameter that determines the shape of the distribution. 

Since $D$ has a factorized form the logarithm in Eq. (\ref{main}) is a sum of functions that depend either on $x$, or $y$, or on $x+y$. Terms with $x+y$ are unimportant because the total energy of a pair is conserved, and finally we obtain the $H$-theorem $H(p')\geq H(p)$ from Eq. (\ref{main}) for
\begin{equation}\label{fin}
H(p)=S(p)+(a-1)\int_0^{\infty}dx\,p(x)\ln x.
\end{equation}
Thus only for the uniform redistribution law, when $a=1$ it is the entropy $S(p)$  that always grows. To some extent this resembles what happens in a system of hard spheres described by the nonlinear Enskog equation, where the $H$-function also differs from the simple Boltzmann entropy \cite{ens}.

This $H$-function is maximized by the equilibrium distribution $p_0(x)\sim x^{a-1}\exp(-\lambda x)$. This solution is already known as equilibrium one for the pure gambling model of Bassetti and Toscani \cite{T}, which is in fact a continuous time version of Ulam's redistribution problem.

There are several ways to understand the result (\ref{fin}). First of all the $H$-function from (\ref{fin}) may be rewritten as $H=-\int dxf\ln(f/x^{a-1})$. This suggests that probably the additional term in $H$ may be related to the `graining' with which the space of $x$ is actually resolved \cite{et}. One may also view $H$  as the usual entropy, but for some multidimensional problem. Indeed, suppose that in $d$ dimensions we have some `velocity' distribution $\phi({\bf v})$. Then if $\phi$ depends only on the absolute value $x=|{\bf v}|$  we may introduce a new function $p(x)\sim x^{d-1}\phi(x)$, normalized as $\int dx p(x)=1$, and the entropy equals
$$
S\sim -\int_0^{\infty} dx\, x^{d-1}\phi\ln\phi\sim-\int_0^{\infty} dx\,p\ln(p/x^{d-1})
$$
up to a constant. So probably Ulam's problem with arbitrary $a$ is a projection of some yet unknown $a$-dimensional problem with uniform redistribution of some vector quantity when only absolute values are taken into account.

Another way to look at the same result is to rewrite the beta redistribution law as
$$
D\left(\frac{x}{x+y}\right)\sim \exp\left[-E(x)-E(y)+2E(x+y)\right],
$$
where $E(x)=-(a-1)\ln x$. This expression looks like a transition probability for a system in contact with some thermal bath, so that exponents are just Boltzmann factors with $E(x)$ playing the role of energy. But for such a system it is the free energy that always decreases. It is easy to see then that $H$-function from Eq. (\ref{fin}) is exactly  minus this free energy, the last term being due to the average of the new `energy' $E$. Thus, though the conservation of the mean energy $\langle x\rangle$ suggests that our initial system is closed, it is possible to view it as an open one with quite different `energy' $E(x)\sim\ln x$ and at a finite temperature.

Finally, one may try to relate the last term in $H$ to  possible entropy production in a device that actually performs energy redistributions and acts  as a kind of Maxwell's demon.

\section{Conclusions}
We were able to derive an $H$-theorem for Ulam's redistribution problem only for a special set of redistribution laws given by symmetric beta distributions. The $H$-function differs from the usual Boltzmann entropy and has an additional term which can be interpreted in different ways (cf. \cite{ens,et}).

But for an arbitrary redistribution law we encounter a difficulty, because now the initially factorized $f(x,y)$ always looses this property after a collision. Therefore while we may have equilibrium solution for $p(x)$, such a solution for $f(x,y)$ is not possible, because in this case e.g. a stationary solution of Eq. (\ref{lin})  does not have the required factorized form in variables $x$ and $y$ and hence is changed under the subsequent reduction (\ref{prime}). Thus in terms of two particle distribution function we probably deal with relaxation not to equilibrium, but rather to some non-equilibrium steady state which makes the search for the $H$-theorem much more difficult.

\end{document}